\PassOptionsToPackage{colorlinks,citecolor=blue}{hyperref}
\documentclass[aps,prd,nofootinbib,showpacs,superscriptaddress,preprint]{revtex4-1}
\pdfoutput=1
\usepackage[utf8]{inputenc}
\usepackage{Baskervaldx}
\usepackage[baskervaldx]{newtxmath}
\usepackage{orcidlink}
\usepackage{xcolor}
\usepackage{slashed}
\usepackage{algorithm2e}
\usepackage{subcaption}    
\usepackage[font=small,labelfont=bf]{caption}  
\usepackage{comment}
\usepackage{graphicx}
\usepackage{multirow}
\usepackage{float}
\DeclareUnicodeCharacter{2212}{-}

\begin{document}

\title{Type I seesaw mechanism at TeV scale \\ or below with minimal fields}

\author{Kunal Pandey \orcidlink{0009-0006-0098-1517}}
\email{kunal@ctp-jamia.res.in}
\affiliation{Centre for Theoretical Physics,\\ Jamia Millia Islamia (Central University), New Delhi - 110025, India}

\author{Rathin Adhikari \orcidlink{0000-0002-8764-9587}}
\email{radhikari@jmi.ac.in}
\affiliation{Centre for Theoretical Physics,\\ Jamia Millia Islamia (Central University), New Delhi - 110025, India}

\begin{abstract}

A novel scenario is presented within the Type-I seesaw mechanism in which no other beyond Standard Model fields except three heavy right handed neutrinos, have been considered. Light neutrino masses around sub eV scale, could be possible at low seesaw scale around  TeV or even below that. 
 At the leading order, $6 \times 6$ seesaw mass matrix reproduces three massless neutrinos.  The Dirac mass matrix with one loop corrections, breaks that massless texture and it is possible to get massive neutrinos. We have obtained the expression of  mixing and Dirac $CP$ violating phase for light neutrino mass matrix with one loop corrections.
 Only unknown parameters are the Yukawa couplings related to right handed neutrinos and their masses. These parameters satisfy the ATLAS, CMS experimental constraints.
\end{abstract}

\maketitle

\section{Introduction} 
Neutrino oscillation data indicates \cite{11} that neutrinos possess a very small but non-zero mass ($\sim 0.1$ eV) and hence the Standard Model (SM) needs to be extended in order to accommodate neutrino masses as well as mixing among different flavors of light neutrinos. The canonical Type-I seesaw \cite{Yanagida:1979as,Gell-Mann:1979vob} mechanism remains to this date the most minimal approach of addressing the mass problem by introducing three heavy Right-handed neutrinos (RHNs) to the particle content of the Standard Model (SM). If one considers the neutrino-Higgs Yukawa couplings to be of the order of tau-Higgs Yukawa coupling then the mass scale of the heavy RHNs (also called the seesaw scale) is found to be $\sim 10^9$ GeV. The experimental verification of such a high mass-scale seems difficult and a natural question arises : Is it possible to lower down such a high scale of seesaw? To address this issue various studies in several directions have been performed like the Extended Double Seesaw model \cite{Kang:2006sn}, Inverse seesaw \cite{Mohapatra:1986aw} and Linear seesaw \cite{Akhmedov:1995ip} but, in all these works the `minimal' nature of the seesaw gets destroyed as one introduces additional fermions (in addition to three RHNs) and/or scalar fields to address the neutrino mass problem while aiming for a reasonably good testable model. So a desirable line of work is then to explore if a lower seesaw scale is achievable with only the field contents of the standard Type-I seesaw so as to increase its testability in colliders. In this spirit there are some works \cite{Ker,Buchmuller:1991tu,Buchmuller:1990du} which introduce some special massless textures for the neutrino Dirac mass matrix ($M_{D}$) in which all the three light neutrinos are mass-less at the tree level while $M_{D}$ remains non-zero. Then the light neutrino masses are obtained via small perturbations but they also require an extended scalar sector \cite{Adhikari:2010yt} or additional fermions \cite{Ker} along with some spurious symmetries which results in some elements of $M_{D}$ or the light neutrino mass matrix ($M_{\nu}$) to become identically zero (called texture-zeros). In this work we have attempted to address these inherent issues while dealing with massless-textures without the addition of more fields or texture zeros to arrive at the correct order of light neutrino masses with a lowered scale of seesaw ($\sim 1$ TeV or below that). The basic idea is that one has to consider one loop corrections to Dirac mass matrix $M_D$ originating from W, Z and Higgs bosons via the heavy RHNs through light-heavy mixing to the massless texture of $M_D$ in order to account for neutrino mass and mixing. These corrections break the massless texture at one loop level. Any earlier work of adding this one loop corrections with minimal fields to $M_D$ for tree level massless light neutrino seesaw texture, is not known to our knowledge \footnote{With non-minimal fields massless texture was found earlier to be broken at the two loop level \cite{Adhikari:2010yt}}.  We have discussed the modified seesaw formula for light neutrino masses after inclusion of one loop corrections. In this scenario, we have obtained the expression of mixing among  different flavors of active neutrinos. In sec-(\ref{sec1}), we have discussed the massless texture of $6 \times 6$ seesaw mass matrix. In sec-(\ref{sec2}), we discuss the conditions on higher order corrections required for breaking massless seesaw texture and then show that the one-loop corrected modified $M_D$ with minimal field content in the model, does break the massless seesaw texture. 
In sec-(\ref{sec3}), after including one loop corrections, we have discussed the modified seesaw formula for light neutrino masses and mixing and Dirac $CP$ violating phase. Lastly, we have discussed  how with the variation of seesaw scale and Yukawa couplings, one may obtain suitable light neutrino masses. These Yukawa couplings and the right handed neutrino masses, are discussed to be well within the experimental limit. 
Finally, we present our concluding remarks in sec-(\ref{sec4}).

\section{Massless Texture of Seesaw Mass Matrix At Tree Level :} \label{sec1}

The standard Type-1 seesaw Lagrangian is considered which requires the addition of only three heavy right-handed Majorana neutrinos $(N_R)$ to the Standard Model particle content along with a bare mass term for the $N_R$ fields:

\begin{equation}
                        \mathcal{L}_Y = -Y_{\imath \jmath}\Bar{L_\imath}\Phi l_{R\jmath} - Y_{\imath \jmath}^{\nu} \Bar{L}_\imath \Tilde{\Phi} N_{R_{\jmath}}  -\frac{1}{2}\Bar{N}^c_{R_{\imath}} M_{R_{\imath \jmath}} N_{R\jmath}+  h.c
                    \end{equation}
where $L_i$ are the lepton doublets, $\Phi$ is the Higgs-doublet and $N_{R_{j}}$ are the heavy Right-handed neutrinos. After spontaneous symmetry breaking, the second term forms the neutrino Dirac mass matrix $M_{D_{ij}} = Y_{ij}^{\nu} v $ (where $v$ is the Higgs vev) and last term forms the heavy Majorana neutrino mass matrix $M_R$ (which, for simplicity, is considered a diagonal matrix).
The full Type-I seesaw mass matrix is a $6\times6$ matrix which consists of four $3\times3$ sub-matrices $M_{D}$, $M_{R}$, $M^{T}_{D}$ and $O$ matrices\cite{11}. This can be written in the basis $(\nu_e, \nu_{\mu}, \nu_{\tau}, N_{R_{1}}, N_{R_{2}}, N_{R_{3}})$ as: 
\begin{equation} \label{eq2}
    \mathcal{M}_{\textit{}{seesaw}}= \begin{pmatrix}
        O & M_{D} \\
        M_{D}^{T} & M_{R}
    \end{pmatrix}
\end{equation}

Owing to the tedious process of diagonalization to attain neutrino masses, seesaw approximation \cite{Grimus:2000vj} is usually applied to get the light neutrino mass matrix which, after diagonalizaion gives the light neutrino masses. These light mass eigenvalues are suppressed by the scale of the heavy right handed neutrino mass which is the seesaw scale. The most general structures for $M_{D}$ and the diagonal $M_{R}$ could be written as:

\begin{equation} \label{eq3}
                    M_D=\begin{pmatrix}
                        x_1  & x_2  & x_3 \\
                        \alpha_1 x_1  & \alpha_2 x_2  & \alpha_3 x_3 \\
                        \beta_1 x_1  & \beta_2 x_2  & \beta_3 x_3 
                    \end{pmatrix}
                \end{equation}
                \begin{equation}
                    M_R=\begin{pmatrix}
                        M_1 & 0 & 0 \\
                        0 & M_2 & 0\\
                        0 & 0 & M_3
                        \end{pmatrix}
                    \end{equation}
where the elements $x_i$, $\alpha_{i}$, $\beta_{i}$ and $M_{i}$ in general, may be complex. 
As shown in  \cite{Ker,Adhikari:2010yt}, at the leading order, for three light neutrinos to be massless, one may consider the following conditions : 

\begin{eqnarray}
\alpha_1 = \alpha_2 = \alpha_3 \;,
  \nonumber \\
\beta_1 = \beta_2 = \beta_3 \; ,
\end{eqnarray}
in $M_{D}$, then we will have two massless neutrinos at the tree level. This amounts to making any two columns of $M_D$ proportional to each other and on the top of this if one further puts the condition:
\begin{equation}
    \label{eq4}
    \frac{x_1^{2}}{M_1}+\frac{x_2^{2}}{M_2} + \frac{x_3^{2}}{M_3} = 0
\end{equation}
then one obtains all the three light neutrinos to be massless at the tree level itself. Such massless-ness for light neutrinos has been extensively studied in the literature \cite{Ker,Buchmuller:1991tu,Buchmuller:1990du}. It would have been nice if these features of the Yukawa couplings could somehow be motivated from symmetry principles but as demonstrated in \cite{He:2009xd} it seems unlikely to be the case. Our focus is more phenomenological rather than a theoretical one where the scaling feature is adopted as kind of a `mathematical theorem' and its usefulness is demonstrated through a setup that provides a mechanism to obtain light neutrino masses for seesaw scale around TeV or below with Yukawa couplings of neutrinos somewhat nearer to other SM Yukawa couplings.
 
 However,  one needs massive light neutrinos to explain the neutrino oscillation data. For that, one loop corrections to $M_D$ is to be considered and is discussed in the next section.  

\section{One-Loop Corrections To $M_{D}$:} \label{sec2}

Higher order corrections to massless texture has been studied earlier \cite{Adhikari:2010yt,Ker} with some extra scalar field and it has been shown that two loop corrections are required to break the massless texture of $M_D$. This is because  their one-loop corrections  come out to be directly proportional to $M_{D_{ij}}$ and hence the massless texture remain intact. In our work, without considering any new scalar field, we have identified the possible  three one-loop Feynman diagrams involving $N_{Rj}$ and other SM fields like W, Z and Higgs boson as shown in Fig.(\ref{fig1}) . These one loop  corrections successfully break the massless texture as  the corrections are  proportional to higher powers of $M_{D_{ij}}$.

 The diagrams involving W and Z bosons via the heavy-light mixing are described by the following gauge interactions:
\begin{equation}
    \begin{aligned}
    - \mathcal{L}_{g} \subset & \frac{g}{\sqrt{2}} W_{\mu}^{+} \sum_{i, j=1}^3 [\Bar{N}_{i}^c (V^{\dagger})_{ij} \gamma^{\mu} P_{L} l_{j} ]  +  \frac{g}{2 C_{\theta_{w}}} Z_{\mu} \\
   & \times \sum_{i, j=1}^3 [\Bar{\nu_{i}} (U^{\dagger} V)_{ij} \gamma^{\mu} P_{L} N_{j}^c + \Bar{N_{i}}^c (V^{\dagger}V)_{ij} \gamma^{\mu} P_{L} N_{j}^c ]
   \end{aligned}
\end{equation}
where $P_{L}= \frac{1- \gamma^5}{2}$, g is the gauge coupling constant, $C_{\theta_{w}} = \cos\theta_{w}$, $\theta_{w}$ is the weak mixing angle, $U$ is the Pontecorvo-Maki-Nakagawa-Sakata (PMNS) matrix and $V_{ij}= \frac{M_{D_{ij}}}{M_{N_{j}}}$ is the heavy-light mixing matrix element through which the heavy right handed-neutrinos interact with the W and Z bosons. The full expressions of the various one-loop corrections ($\epsilon^{W}$, $\epsilon^{Z}$ and $\epsilon^{H}$) are given in the Appendix(\ref{Appendix:A}) in which contributions from three one loop Feynman diagrams, have been summed over as:
\begin{equation}
    \epsilon_{ij} = \epsilon_{ij}^{W} + \epsilon_{ij}^{Z} + \epsilon_{ij}^{H}
\end{equation}

Next, we discuss the condition on $\epsilon_{ij}$ for which the massless texture of $M_D$ will be broken. The one-loop corrected Dirac mass matrix is given by: 

\begin{figure}[h!]
  \centering
  \begin{subfigure}[b]{0.4\textwidth}
    \centering
    \includegraphics[width=\textwidth]{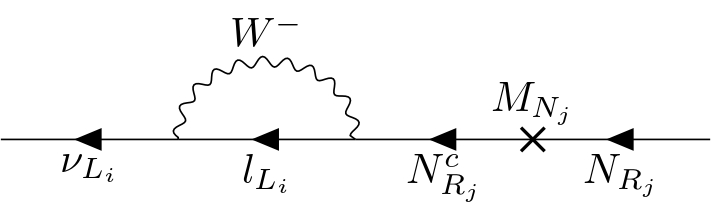}
    \caption{W boson correction}
    \label{fig:subfig1}
  \end{subfigure}
  \hfill
  \begin{subfigure}[b]{0.4\textwidth}
    \centering
    \includegraphics[width=\textwidth]{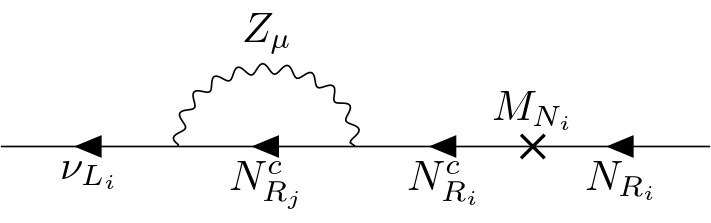}
    \caption{Z boson correction}
    \label{fig:subfig2}
  \end{subfigure}

  \vspace{0.5cm}  
  \begin{subfigure}[b]{0.4\textwidth}
    \centering
    \includegraphics[width=\textwidth]{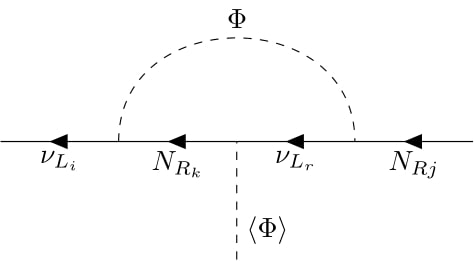}
    \caption{Higgs boson correction}
    \label{fig:subfig3}
  \end{subfigure}
  
  \caption{One-loop corrections to the Dirac Mass matrix ($M_D$)}
  \label{fig1}
\end{figure}

\begin{equation}
                    M_{D}'=\begin{pmatrix}
                        x_1 + \epsilon_{11}   & x_2 + \epsilon_{12}  & x_3  + \epsilon_{13}\\
                        \alpha x_1 + \epsilon_{21}  & \alpha x_2 + \epsilon_{22}  & \alpha x_3  + \epsilon_{23}\\
                        \beta x_1 + \epsilon_{31}  & \beta x_2 + \epsilon_{23}  & \beta x_3   + \epsilon_{33}
                    \end{pmatrix}
\end{equation}
We have to check whether after including such corrections, one can write  $M_{D}^{'}$ again in the form of $M_D$ with some transformation of $x_i$ to $x_{i}^{'}$ and $\alpha_i$ and $\beta_i$ to  $\alpha_{i}^{'}$ and $\beta_{i}^{'}$ . If in such a way  $M_{D}^{'}$ can not be written, then the massless texture in $M_{D}^{'}$ will be broken.
Let us define the new parameters as:
\begin{align}
    x_{1}'=x_1 + \epsilon_{11} \\
    x_{2}'=x_2 + \epsilon_{12} \\
    x_{3}'=x_3 + \epsilon_{13} \\
\end{align}
Then the new $\alpha$'s and $\beta$'s become:
\begin{align}
    \alpha_{1}'= \alpha + \frac{\epsilon_{21} - \alpha \epsilon_{11}}{x_{1}'} \\
    \beta_{1}'= \beta + \frac{\epsilon_{31} - \beta \epsilon_{11}}{x_{1}'} \\
    \alpha_{2}'= \alpha + \frac{\epsilon_{22} - \alpha \epsilon_{12}}{x_{2}'} \\
    \beta_{2}'= \beta + \frac{\epsilon_{32} - \beta \epsilon_{12}}{x_{2}'} \\
    \alpha_{3}'= \alpha + \frac{\epsilon_{23} - \alpha \epsilon_{13}}{x_{3}'} \\
    \beta_{3}'= \beta + \frac{\epsilon_{33} - \beta \epsilon_{13}}{x_{3}'} 
\end{align}
These new $\alpha's$ and $\beta's$ should strictly differ from each other  in order to break the scaling feature. To keep track of this and also of the differences between the new scaling factors we define:
\begin{align}
    \alpha_{i}' - \alpha_{1}'  = \Delta \alpha_{i1}' = \frac{x_{1}' \epsilon_{2i} - x_{i}' \epsilon_{21} + \alpha (\epsilon_{11} x_{i}' - \epsilon_{1i} x_{1}')}{x_{1}' x_{i}'} \\
     \beta_{i}' - \beta_{1}' =  \Delta \beta_{i1}' = \frac{x_{1}' \epsilon_{3i} - x_{i}' \epsilon_{31} + \beta (\epsilon_{11} x_{i}' - \epsilon_{1i} x_{1}')}{x_{1}' x_{i}'}
\end{align}
where $i =2,3$ and $\Delta \alpha_{i1} = \alpha_{i} - \alpha_{1} $, $ \Delta \beta_{i1} = \beta_{i} - \beta_{1}$ and for the scaling to be broken we must have  $\Delta \alpha_{i1} \neq 0$ and $\Delta \beta_{i1} \neq 0$ for two light neutrinos to be massive. In some other models with more scalars, the one-loop corrections may not change the massless texture. The signature of the change in the massless texture due to higher order corrections is given by non-zero values of  $\Delta \alpha_{i1}$ and $ \Delta \beta_{i1}$. In some models with more scalar fields sometimes it is found that the one-loop correction doesn't make this $\Delta \alpha_{i1}$ and $ \Delta \beta_{i1}$ to be non-zero. Then, one has to consider further higher order corrections at two-loop level. But, in our case with no further addition to the scalar sector we have found that $\Delta \alpha_{i1}$ and $ \Delta \beta_{i1}$ to be non-zero (which can be verified by considering $\epsilon_{ij}$ as given in Appendix(\ref{Appendix:A})) thus, breaking the massless texture. So, in our case, the one-loop correction gives non-zero light neutrino masses. It is to be noted that even after satisfying above conditions, in the most minimal scenario one massless neutrino may be  obtained due to condition mentioned in Eq.(\ref{eq4}).  In principle, this could also be a viable scenario as only the mass-squared differences of neutrino masses are obtained from neutrino oscillation experiments and as a result one neutrino could indeed be massless. For instance, considering $x_1 \simeq 0.1$ GeV, $x_2 \simeq 0.2$ GeV and $x_3 \simeq 0.7$ GeV along with $\alpha = 1$ and $\beta = 1.1$, while the heavy-right handed neutrino masses $M_{1}$, $M_{2}$ and $M_{3}$ as $1$ TeV, $2.5$ TeV and $18$ TeV respectively in accordance with Eq.(\ref{eq4}),  the differences in the new scale factors (which are actually responsible for neutrino mass generation) are obtained as : $\Delta \alpha_{21} \simeq 5 \times 10^{-11}$, $\Delta \alpha_{31} \simeq 10^{-7}$, $\Delta \beta_{21} \simeq 2 \times 10^{-10}$ and $\Delta \beta_{31} \simeq 5 \times 10^{-7}$. As  shown in the next section, these $\epsilon_{ij}$ corrections will play a vital role in neutrino mass generation while realizing a low scale of seesaw.

\section{Light Neutrino Masses and Mixings: } \label{sec3}
Another crucial outcome of the massless conditions is that if one calculates the light neutrino mass matrix using the seesaw formula then it comes out to be a zero matrix as shown below:
\begin{equation}
    \label{eq22}
  \mathcal{M}_\nu = -M_D M_R^{-1} M_D^{T} \equiv 0
\end{equation}
Now, after one loop corrections, $M_{D}$ gets modified and if one collectively treats   $\epsilon_{ij}$ as corrections and corresponding  correction matrix as $\mathcal{E}$, then  we can write:

\begin{equation}
  \mathcal{M}_{\nu}' = -(M_D + \mathcal{E})M_R^{-1} (M_D + \mathcal{E})^{T}
\end{equation}
Using  Eq.(\ref{eq22}) and ignoring higher order in $\mathcal{E}$, one can write above equation as
\begin{eqnarray}\label{eq25}
    \mathcal{M}_{\nu}' \approx -\mathcal{E} M_{R}^{-1} M_{D}^{T} - M_{D} M_{R}^{-1} \mathcal{E}^{T}            \end{eqnarray}
Because of the massless texture  of light neutrino masses at the tree level, Eq.(\ref{eq25}) may be considered as the modified seesaw formula for massive light neutrino mass matrix. 
The correct order of light neutrino masses could be accommodated at a much lower scale of $M_{R}$, for instance, for $\epsilon \sim 10^{-7}$ GeV and $M_{D_{ij}} \sim 0.1$ GeV one can obtain the correct order for light neutrino masses around 0.1 eV  for  TeV scale of $M_{R}$.

It is also possible to account for the mixing angles too in the above setup. The  light neutrino mass matrix obtained in Eq.(\ref{eq25}) is not in general, hermitian and to get the masses and mixing (PMNS matrix), one works with the following Hermitian matrix:
\begin{equation}
    \mathcal{M}=\mathcal{M}_{\nu}^{\dagger '} \mathcal{M}_{\nu}'
\end{equation}
This procedure results in the loss of two complex phases which are actually Majorana phases and we are left with only one $CP$ -violating Dirac phase. Using Eq.(\ref{eq25}), one starts with the one-loop corrected light neutrino mass matrix as shown in the Appendix(\ref{AppB}) and then following the procedure outlined in \cite{Aizawa},  one can first determine  $\tan\theta_{23}$ using their Eq. (20) and similarly, other mixing angles and $CP$ violating Dirac phase ( some of which are expressed in terms of $\theta_{23}$) could be obtained.  In terms of different elements of matrix $\mathcal{M}_{\nu}'$ (which are explicitly shown in Appendix(\ref{AppB})), one can write different mixing angles as shown below: 
\begin{equation}
    \tan \theta_{23}=\frac{Im(Q)}{Im(R)}
\end{equation}

\begin{equation}
    \tan2\theta_{12}= \frac{2 \eta}{K_2-K_1}
\end{equation}

\begin{equation}
    \tan\delta= \frac{-Im(Q)}{s_{23}(s_{23}Re(Q)+c_{23}Re(R))}
\end{equation}

\begin{equation}
    \tan2\theta_{13}=\frac{2 |s_{23}Q+c_{23}R|}{K_3-P}
\end{equation}
where $s_{ij} = \sin \theta_{ij}$ and $c_{ij} = \cos \theta_{ij}$ and for simplicity, their experimental values will be used on the right hand side of the above equations.
In order to determine the $CP$-violating phase one needs a complex parameter and so $\alpha$ ($=\alpha_r + i \alpha_i$) has been promoted to a complex number and all other parameters in $M_D$ in Eq.(\ref{eq2}) have been assumed to be real. For giving two sets of benchmark values, as examples, we have chosen $ \beta \sim -1.88 \; , \; \alpha_{r} \sim 0.33 \; , \; \alpha_{i} \sim -8 \;. $
 
In Table(\ref{tab1}), we have shown two sets of benchmark values 
which gives three light neutrino masses as $ m_1 \sim 0 \; \mbox{eV} ,\; m_2 \sim 0.0034 \; \mbox{eV},\; m_3 \sim 0.035 \; \mbox{eV} $ for BP1 and 
$ m_1 \sim 0 \;\mbox{eV} ,\; m_2 \sim 0.0018 \;\mbox{eV},\; m_3 \sim 0.034 \;\mbox{eV} $ for BP2.
\begin{table}[h!]
\centering
\begin{tabular}{|c|c|c|c|c|c|c|}
\hline
\textbf{Benchmark Point} & $x_1$ (GeV) & $x_2$ (GeV) & $x_3$ (GeV) & $M_1$ (GeV) & $M_2$ (GeV) & $M_3$ (GeV) \\ \hline
\textbf{BP1} & 0.00246 & 0.0123 & 0.0246 & -40 & -1000 & 2000 \\ \hline
\textbf{BP2} & 0.00492 & 0.00615 & 0.0123 & -64 & -100 & 200 \\ \hline
\end{tabular}
\caption{Parameter values for two benchmark points (BP1 and BP2).} \label{tab1}
\label{tab:bp_parameters}
\end{table}
It is possible to consider the variations of such benchmark values easily and consider further lowering of the seesaw scale in the following way. The Yukawa couplings $Y_{ij}^\nu$ and $M$ are constrained by Eq.(\ref{eq4}) which one could also write in terms of some transformed $ x_i^{''}  = \gamma_1 x_i$ and $M_i^{''}= \gamma_2 M_i$ then still Eq.(\ref{eq4}) is satisfied for transformed parameters as
\begin{eqnarray} \label{eq30}
 \frac{{x_1^{''}}^{2}}{M_1^{''}}+\frac{{x_2^{''}}^{2}}{M_2^{''}} + \frac{{x_3^{''}}^{2}}{M_3^{''}} = \gamma \left\{\frac{x_1^{2}}{M_1}+\frac{x_2^{2}}{M_2} + \frac{x_3^{2}}{M_3} \right\} =  0
\end{eqnarray}
where $\gamma = \gamma_1^2/\gamma_2$ is chosen to be some constant value.  Such transformations will maintain the massless texture of three light neutrinos at the tree level but with one loop corrections two neutrinos will be massive. We discuss below, how the light neutrino masses are changed due to such transformation considering transformed one lop corrections $\epsilon_{ij}^{''}$ by considering the transformed $x_i^{''}$ and $M_i^{''}$ in place of $x_i$ and $M_i$ in $\epsilon_{ij}$ shown in the Appendix(\ref{Appendix:A}).

In Fig(\ref{fig2}) and Fig(\ref{fig3}), we have considered the transformation of benchmark points in BP1 and BP2 respectively. In Figures we have shown the variation of $\gamma_2$ so that $\gamma_1 = \sqrt{\gamma_2}$ which implies $\gamma=1$. Considering the transformations of both $x_i$ and $M_i$ mentioned in BP1 and BP2, we have shown the variations of light neutrino masses $m_2$ and $m_3$ to be small. This is due to some pattern of ${M_D}_{ij}$ and $M_j$ present in $\epsilon_{ij}$ in Appendix(\ref{Appendix:A}). However, $m_1$ remains massless. Considering $M_i^{''}= \gamma_2 M_i$, the variation of $\gamma_2$ is shown in Fig(\ref{fig2}) and benchmark values are given in Table(\ref{tab1}).  One can see that the appropriate light neutrino masses could be obtained at different seesaw scale. As for example, corresponding to $M_i$ in BP2, which are already around the electroweak scale, if we consider $\gamma_2$ to be $0.2$, then the transformed heavy right handed neutrino masses will be $1/5$th of $M_i$ shown in the Table(\ref{tab1}) for BP2. So heavy right handed neutrino masses could be further lowered. For such transformed seesaw scale, the light neutrino masses, does not differ significantly and have been shown in Fig(\ref{fig3}).

\begin{figure}[htbp]
    \centering
    \begin{subfigure}{0.45\textwidth}
        \centering
        \includegraphics[width=\textwidth]{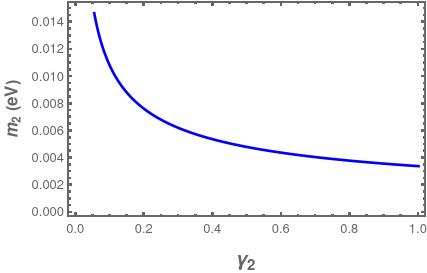} 
        \caption{Variation of $m_{2}$ with $\gamma_2$.}
        \label{fig:subfig1}
    \end{subfigure}
    \hfill
    \begin{subfigure}{0.45\textwidth}
        \centering
        \includegraphics[width=\textwidth]{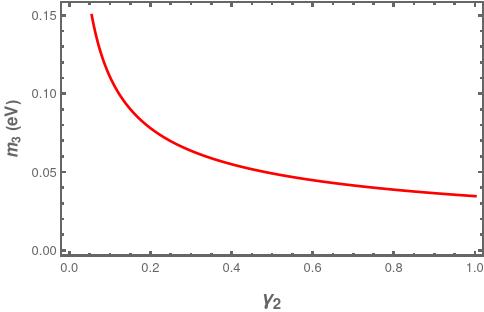} 
        \caption{Variation of $m_{3}$ with $\gamma_2$.}
        \label{fig:subfig2}
    \end{subfigure}
    \caption{Plots for BP1: considering change in both $x_i$ and $M_i$ with $\gamma_1 = \sqrt{\gamma_2}$ implying $\gamma =1$.} 
    \label{fig2}
\end{figure}

\begin{figure}[htbp]
    \centering
    \begin{subfigure}{0.45\textwidth}
        \centering
        \includegraphics[width=\textwidth]{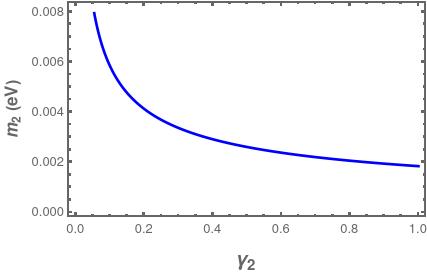} 
        \caption{Variation of $m_{2}$ with $\gamma_2$ .}
        \label{fig:subfig1}
    \end{subfigure}
    \hfill
    \begin{subfigure}{0.45\textwidth}
        \centering
        \includegraphics[width=\textwidth]{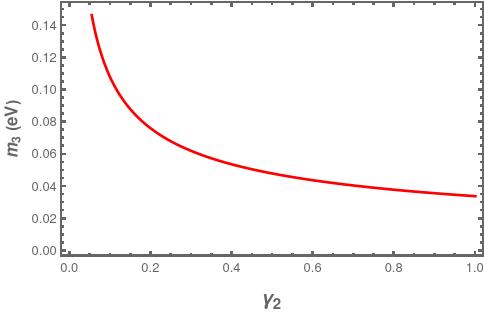} 
        \caption{Variation of $m_{3}$ with $\gamma_2$.}
        \label{fig:subfig2}
    \end{subfigure}
    \caption{Plots for BP2: considering change in both $x_i$ and $M_i$ with $\gamma_1 = \sqrt{\gamma_2}$ implying $\gamma =1$}
    \label{fig3}
\end{figure}

Some region of heavy to light neutrino mixing versus the right handed neutrino mass has been disallowed   from different CMS, ATLAS experimental constraint \cite{Abada:2022wvh} on collider dilepton, trilepton, and long-lived searches for heavy neutral lepton which is heavy right-handed neutrino in our case.  Considering the relation of mixing with the Yukawa couplings as $|U_{\alpha i}|^2 \sim |Y_{ij} v_u /M_j|^2$ \cite{Drewes:2019mhg} for diagonal $M_R$ in the seesaw mass matrix, 
we have interpreted the most stringent experimental bounds on mixing versus $M_j$
in terms of  the Yukawa coupling $Y_{ij}$ versus $M_j$ . Considering $\alpha$ and $\beta$ of the order of 1 in Eq.(\ref{eq3}), the Yukawa couplings can be written as $Y_{ij}= M_{D_{ij}}/v \approx x_j/v \approx Y_j$  and the corresponding right handed neutrino mass is $M_j$. Experimentally, for right handed neutrino mass $M_j$ around 500 GeV, 100 GeV, 50 GeV, 10 GeV, the upper bound of Yukawa couplings are constrained about 0.9, 0.1, 0.001, $5 \times 10^{-5}$. For the benchmark values considered in BP1 and BP2, as well as for the variations of the benchmark values, the above-mentioned experimental constraints are found to be satisfied. As for example, even for the heavy right handed neutrino mass $M_{1} \sim -64 $ GeV mentioned in BP2 Table(\ref{tab1}), the corresponding Yukawa coupling is $ x_1/v \sim Y_1  \sim 2 \times 10^{-5}$. When these $Y_1$'s and $M_{1}$'s are simultaneously varied as per Eq(\ref{eq30}), then even for $M^{''}_{1} = \gamma_2 M_1 \sim -12.8$ GeV and  the Yukawa coupling is $ Y_1^{''} = \sqrt{\gamma_2} Y_1 \sim 4.02 \times 10^{-6}$  for $\gamma_2 \sim 0.201$ giving the correct order of the light neutrino masses (see Fig(\ref{fig3})) while simultaneously satisfying the above stated experimental bounds. 

\section{Conclusion} \label{sec4}

With minimal fields, with only three right handed neutrinos as beyond SM fields, we have discussed, how seesaw scale could be around TeV to electroweak scale. With the massless texture at the tree level, the modified seesaw formula for light neutrino mass matrix has been given in Eq.(\ref{eq25}). The details of the the one loop corrections, possible in this scenario, has been given in Appendix(\ref{Appendix:A}). The expressions of mixing angles and Dirac $CP$ violating phase, have been obtained after taking into account one loop corrections. Only unknown parameters are the Yukawa couplings for the heavy right handed neutrinos and their masses which are presently not constrained  by CMS and ATLAS experiments for their values considered in this work but it could be possible to constrain in near future.

\hspace*{\fill}

\textbf{Acknowledgement}: Kunal Pandey would like to thank Imtiyaz Ahmad Bhat for useful discussions and Subhodeep Sarkar for initial help with computations.

\appendix

\section{One-Loop Corrections:} \label{Appendix:A}
The one-loop corrections to $M_D$ involving Higgs, W and Z bosons as internal line in different Feynman diagrams, are given below as
\begin{equation} \label{eqa1}
\epsilon_{ij} = \epsilon^W_{ij} + \epsilon^Z_{ij} + \epsilon^H_{ij}
\end{equation}
where,
\begin{align}
\epsilon_{ij}^{W} &= \frac{1}{32 \pi^2 M_{D_{ij}}^3 M_{N_i} \left( M_{D_{ij}} - M_{N_j} \right)} g^2 M_{D_{ji}} M_{N_j} \\ \nonumber
&\times \left[ -M_{D_{ij}}^2 \left( -2 M_W^2 + 3 M_{D_{ij}}^2 + 2 m_{l_i}^2 \right) 
- 2 M_{D_{ij}}^4 \log \left( \frac{\mu^2}{m_{l_i}^2} \right) \right. \\ \nonumber
&\left. - \left( 2 M_W^2 M_{D_{ij}}^2 + M_{D_{ij}}^4 - \left( M_W^2 - m_{l_i}^2 \right)^2 \right) 
\log \left( \frac{m_{l_i}^2}{M_W^2} \right) \right. \\ \nonumber
&\left. - 2 \left( -M_W^2 + M_{D_{ij}}^2 + m_{l_i}^2 \right) f_{ij} 
\log \left( \frac{1}{2 M_W m_{l_i}} \left( M_W^2 - M_{D_{ij}}^2 + m_{l_i}^2 + f_{ij} \right) \right) \right],
\end{align}
where we have defined:
\begin{align}
f_{ij} &= \sqrt{ M_{D_{ij}}^4 + \left( M_W^2 - m_{l_i}^2 \right)^2 
- 2 M_{D_{ij}}^2 \left( M_W^2 + m_{l_i}^2 \right)}.
\end{align}
\begin{align}
\epsilon_{ij}^{Z} &= -\frac{g^2  \, M_{D_{ji}}^2}{64 \, C_{w\theta}^2 \, \pi^2 \, M_{D_{ij}}^2 \left( M_{D_{ij}} - M_{N_{i}} \right) M_{N_{j}}} \\ \nonumber
&\quad \times \Bigg[ M_{D_{ij}}^2 \left( -2 M_{Z}^2 + 3 M_{D_{ij}}^2 - 6 M_{D_{ij}} M_{N_{j}} + 2 M_{N_{j}}^2 \right) \\ \nonumber
&\quad + 2 M_{D_{ij}}^3 \left( M_{D_{ij}} - 2 M_{N_{j}} \right) \log \left( \frac{\mu^2}{M_{N_{j}}^2} \right) \\ \nonumber
&\quad + \left( 2 M_{Z}^2 M_{D_{ij}}^2 + M_{D_{ij}}^4 - 2 M_{D_{ij}}^3 M_{N_{j}} - \left( -M_{Z}^2 + M_{N_{j}}^2 \right)^2 \right. \\ \nonumber
&\quad\quad \left. + 2 M_{D_{ij}} \left( -M_{Z}^2 M_{N_{j}} + M_{N_{j}}^3 \right) \right) \log \left( \frac{M_{N_{j}}^2}{M_{W}^2} \right) \\ \nonumber
&\quad + 2 \left( -M_{Z}^2 + M_{D_{ij}}^2 - 2 M_{D_{ij}} M_{N_{j}} + M_{N_{j}}^2 \right) \\ \nonumber
&\quad\quad \times \sqrt{ M_{D_{ij}}^4 + \left( -M_{Z}^2 + M_{N_{j}}^2 \right)^2 - 2 M_{D_{ij}}^2 \left( M_{Z}^2 + M_{N_{j}}^2 \right) } \\ \nonumber
&\quad\quad \times \log \Bigg( \frac{M_{Z}^2 - M_{D_{ij}}^2 + M_{N_{j}}^2 + \sqrt{ M_{D_{ij}}^4 + \left( -M_{Z}^2 + M_{N_{j}}^2 \right)^2 - 2 M_{D_{ij}}^2 \left( M_{Z}^2 + M_{N_{j}}^2 \right) }}{2 M_{Z} M_{N_{j}}} \Bigg) \Bigg].
\end{align}

\begin{align}
\epsilon_{ij}^{H} &= \frac{1}{64 \pi^2 M_{D_{ij}}^3 M_{N_{k}}} \, v_h \, Y_{ik} \, Y_{kr} \, Y_{rj} \\ \nonumber
&\times \Bigg[ 2 M_{D_{ij}}^2 M_{N_{k}} \left( 4 M_{D_{ij}} + M_{N_{k}} \right) 
- 2 \left( M_{D_{ij}}^2 - M_{h}^2 \right)^2 \log \left( \frac{M_{h}^2}{-M_{D_{ij}}^2 + M_{h}^2} \right) \\ \nonumber
&\quad + \left( M_{D_{ij}}^4 - 2 M_{D_{ij}}^2 M_{h}^2 + 2 M_{D_{ij}}^3 M_{N_{k}} + \left( M_{h}^2 - M_{N_{k}}^2 \right)^2 \right. \\ \nonumber
&\quad\quad \left. + M_{D_{ij}} \left( -2 M_{h}^2 M_{N_{k}} + 2 M_{N_{k}}^3 \right) \right) \log \left( \frac{M_{h}^2}{M_{N_{k}}^2} \right) \\ \nonumber
&\quad + 2 \left( M_{D_{ij}}^2 - M_{h}^2 + 2 M_{D_{ij}} M_{N_{k}} + M_{N_{k}}^2 \right) \\ \nonumber
&\quad\quad \times \sqrt{ M_{D_{ij}}^4 + \left( M_{h}^2 - M_{N_{k}}^2 \right)^2 - 2 M_{D_{ij}}^2 \left( M_{h}^2 + M_{N_{k}}^2 \right) } \\ \nonumber
&\quad\quad \times \log \Bigg( \frac{-M_{D_{ij}}^2 + M_{h}^2 + M_{N_{k}}^2 + \sqrt{ M_{D_{ij}}^4 + \left( M_{h}^2 - M_{N_{k}}^2 \right)^2 - 2 M_{D_{ij}}^2 \left( M_{h}^2 + M_{N_{k}}^2 \right) }}{2 M_{h} M_{N_{k}}} \Bigg) \\ \nonumber
&\quad + 4 M_{D_{ij}}^3 M_{N_{k}} \log \left( \frac{\mu^2}{M_{h}^2} \right) \Bigg].
\end{align}
where $M_{W}$, $m_{l_{i}}$ are the masses of W boson and charged lepton with $i=1,2,3$ for the mass of electron, muon and tau respectively, while $M_{D_{ij}}$ are the elements of the original massless texture of the Dirac mass matrix ($M_{D}$) and $M_{ij}$ are the matrix elements of the Majorana mass matrix. We point out that we are working with a diagonal Majorana matrix so ultimately the indices of $M_{ij}$ will be equal in numerical calculations.

\section{Neutrino Mixing Details:} \label{AppB}
Using Eq.(\ref{eq25}), we write down the one loop corrected mass matrix as
\begin{equation} \label{b1}
M_{\nu}'=\begin{pmatrix}
a  & b& c \\
 b & d  & e\\
 c  & e  & f
\end{pmatrix}
\end{equation}
where,
\begin{equation}
   a= -\frac{2 \epsilon_{11} x_1}{M_1}-\frac{2 \epsilon_{12}  x_2}{M_2}-\frac{2 \epsilon_{13} x_3}{M_3}
\end{equation}
\begin{equation}
    b = -\frac{\epsilon_{11} x_1 (  \alpha_r + i   \alpha_i)}{M_1}
        - \frac{\epsilon_{12} x_2 (  \alpha_r + i   \alpha_i)}{M_2}
        - \frac{\epsilon_{13} x_3 (  \alpha_r + i   \alpha_i)}{M_3}
        - \frac{\epsilon_{21} x_1}{M_1}
        - \frac{\epsilon_{22} x_2}{M_2}
        - \frac{\epsilon_{23} x_3}{M_3}
\end{equation}
\begin{equation}
    c = -\frac{
          \beta \left[
            \epsilon_{11} M_2 M_3 x_1 
            + M_1 \left( \epsilon_{12} M_3 x_2 + \epsilon_{13} M_2 x_3 \right)
        \right]
        + \epsilon_{31} M_2 M_3 x_1 
        + M_1 \left( \epsilon_{32} M_3 x_2 + \epsilon_{33} M_2 x_3 \right)
    }{
        M_1 M_2 M_3
    }
\end{equation}
\begin{equation}
    d=-\frac{2 i (   \alpha_i-i  \alpha_r)( \epsilon_{21} M_2 M_3 x_1+ M_1 (\epsilon_{22} M_3 x_2+ \epsilon_{23} M_2 x_3))}{ M_1 M_2 M_3}
\end{equation}
\begin{eqnarray}
    e &=& -\frac{
          \beta \left[
            \epsilon_{21} M_2 M_3 x_1 
            + M_1 \left( \epsilon_{22} M_3 x_2 + \epsilon_{23} M_2 x_3 \right)
        \right]
        + (  \alpha_r + i   \alpha_i) \left[
            \epsilon_{31} M_2 M_3 x_1 
            + M_1 \left( \epsilon_{32} M_3 x_2 + \epsilon_{33} M_2 x_3 \right)
        \right]
    }{
        M_1 M_2 M_3
    }\nonumber \\
\end{eqnarray}
\begin{equation} \label{f}
    f=2  \beta \left(-\frac{\epsilon_{31} x_1}{M_1}-\frac{\epsilon_{32}x_2}{M_2}-\frac{\epsilon_{33} x_3}{ M_3}\right)
\end{equation}
in which $\epsilon_{ij}$ are given by Eq(\ref{eqa1}). Using Eq(\ref{b1})-(\ref{f}), one can write
\begin{equation}
M = M_{\nu}'^{\dagger} M_{\nu}'=\begin{pmatrix}
P  & Q& R \\
 Q^{*} & S  & T\\
 R^{*}  & T^{*}  & U
\end{pmatrix}
\end{equation}
where we have defined:
\begin{eqnarray}
    P = |a|^2+|b|^2+|c|^2
\end{eqnarray}
\begin{equation}
    Q=a^{*}b+b^{*}d+c^{*}e
\end{equation}
\begin{equation}
    R =a^{*}c+b^{*}e+c^{*}f
\end{equation}
\begin{equation}
    S=|b|^2+|d|^2+|e|^2
\end{equation}
\begin{equation}
    T=b^{*}c+d^{*}e+e^{*}f
\end{equation}
\begin{equation}
    U=|c|^2+|e|^2+|f|^2
\end{equation}
\begin{equation}
    K_1= c_{13}^{2} P - c_{13} (\Tilde{s_{13}}(s_{23}Q + c_{23}R)+ \Tilde{s_{13}^{*}}(s_{23} Q^{*}+c_{23}C^{*}))+s_{13}^{2} K_3
\end{equation}
where $\Tilde{s_{13}}=s_{13} e^{\iota\delta}$ and $K_2$, $K_3$ and $\eta$ are given by:

\begin{equation}
    K_2=c_{23}^{2} S + s_{23}^2 U - 2 s_{23}c_{23} Re(T)
\end{equation}
\begin{equation}
    K_3= s_{23}^{2}S + c_{23}^2 U+ 2 s_{23} c_{23} Re(T)
\end{equation}

\begin{equation}
    \eta = \frac{c_{23}Re(Q)-s_{23}Re(R)}{c_{13}}
\end{equation}

\hspace*{\fill}

\end{document}